\documentclass[aps,prb,twocolumn,superscriptaddress,floatfix,amsmath,amssymb,longbibliography]{revtex4-2}
\usepackage{amsmath}
\usepackage{amssymb}
\usepackage{braket}
\usepackage{graphicx}
\usepackage[table]{xcolor}
\usepackage{verbatim}
\usepackage{float}
\usepackage{color}
\usepackage{siunitx}
\usepackage{gensymb}
\usepackage{bm}
\usepackage{xcolor}
\usepackage{multirow}
\usepackage{footnote}
\usepackage{bbm}
\usepackage{enumerate}

\usepackage{tikz}
\usepackage{physics}
\usepackage{dsfont}
\usepackage{comment}
\usepackage{enumitem}

\usepackage{CJK}

\usepackage[bottom]{footmisc}

\usepackage{mathrsfs}
\usepackage[scr=rsfs,cal=boondox]{mathalfa}

\begin{document}
\begin{CJK*}{UTF8}{bsmi}
\title{Substrate, temperature and magnetic field dependence of electric polarisation in mixed-stacking tetralayer graphenes}

\author{Patrick Johansen Sarsfield}
\email{patrick.sarsfield@manchester.ac.uk}
\affiliation{Department of Physics and Astronomy, University of Manchester, Oxford Road, Manchester, M13 9PL, United Kingdom}
\affiliation{National Graphene Institute, University of Manchester, Booth St.\ E., Manchester, M13 9PL, United Kingdom}
\author{Aitor Garcia-Ruiz (艾飛宇)}
\email{aitor.garcia-ruiz@phys.ncku.edu.tw}
\affiliation{Department of Physics and Astronomy, University of Manchester, Oxford Road, Manchester, M13 9PL, United Kingdom}
\affiliation{National Graphene Institute, University of Manchester, Booth St.\ E., Manchester, M13 9PL, United Kingdom}
\affiliation{ Department of Physics and Center of Quantum Frontiers of Reasearch and Technology (QFort), National Cheng Kung University, Tainan, 70101, Taiwan}
\author{Vladimir I. Fal'ko}
\affiliation{Department of Physics and Astronomy, University of Manchester, Oxford Road, Manchester, M13 9PL, United Kingdom}
\affiliation{National Graphene Institute, University of Manchester, Booth St.\ E., Manchester, M13 9PL, United Kingdom}

\begin{abstract}
Polytypes of tetralayer graphene (TLG: Bernal, rhombohedral and mixed stacking) are crystalline structures with different symmetries. Among those, mixed-stacking tetralayers lack inversion symmetry, which allows for intrinsic spontaneous out-of-plane electrical polarisation, inverted in the mirror-image pair, ABCB and ABAC stackings. Here, we compare the intrinsic polarisation of such TLGs with the symmetry-breaking effect of a substrate, which can also generate out-of-plane electric dipole moments with different sizes in all four polytypes, including ABCB and ABAC twins. We analyse their temperature and magnetic field dependence, in view of understanding the origin of the recently measured Kelvin probe force microscopy maps of tetralayer flakes \cite{atri_spontaneous_2023}, and notice that the intrinsic contribution could be singled out based on magnetic field dependence of polarisation measured at low temperatures.
\end{abstract}

\maketitle

\end{CJK*}

\section{Introduction}

In the recent years, few-layer graphenes have become a playground for studying different phases of electronic matter, including superconductivity \cite{Cao_Correlated_2018,Cao_Unconventional_2018,zhou_superconductivity_2021}, correlated-insulator \cite{shen_correlated_2020,xu_tunable_2021} and ferromagnetism \cite{wang_room-temperature_2009,pixley_ferromagnetism_2019,sinha_ferromagnetism_2021,seiler_quantum_2022,zhou_isospin_2022,de_la_barrera_cascade_2022,stepanov_quantum_2019,chen_gate-tunable_2023,guerrero-aviles_rhombohedral_2022}. While most of those studies were focused on bilayer graphene (AB) \cite{seiler_quantum_2022,zhou_isospin_2022,de_la_barrera_cascade_2022} or ABA (Bernal) \cite{stepanov_quantum_2019,chen_gate-tunable_2023} and ABC (rhombohedral) \cite{zhou_superconductivity_2021,guerrero-aviles_rhombohedral_2022} trilayers, there is a much broader variety of atomically thin graphitic films with electronic properties sensitive to a particular stacking order
\cite{freise_structure_1962,bernal_structure_1924,McClure_Band_1957,McClure_Theory_1960,mcellistrim_spectroscopic_2023}.
The thinnest graphitic film where this variety can unfold into structures with mixed -- part Bernal and part rhombohedral -- stacking order are tetralayer (TLG) twins ABCB and ABAC, whose crystalline structures represent a mirror image of each other. 
\cite{shi_tunable_2018,wirth_experimental_2022,fischer_spin_2024,garcia-ruiz_mixed-stacking_2023,fischer_spin_2024,garcia-ruiz_mixed-stacking_2023}. These twin structures lack inversion symmetry which, in principle, allows for the intrinsic out of plane electrical polarisation of un-doped TLGs \cite{garcia-ruiz_mixed-stacking_2023}.

An experimental observation of such a polarisation has been recently reported \cite{atri_spontaneous_2023}, based on Kelvin probe force microscopy (KPFM) mapping of tetralayer flakes with various stacking areas, which interpreted the difference of potential above ABCB and ABAC twins as a result of such intrinsic polarisation. However, a surface potential caused by a substrate on the bottom layer of a TLG also can induce an out-of-plane polarisation that would also differ for the two twin structures. Such an influence would be unavoidable in KPFM scanning, as a graphene flake would have to be supported by a substrate with its top surface exposed to a probe. Here, we analyse the substrate, temperature, and magnetic field dependence of electric polarisation of various TLGs aiming to disentangle the substrate-induced ``extrinsic" contributions from the intrinsic inversion asymmetry effect specific for ABCB/ABAC twins.

The analysis presented in this paper is based on a self-consistent modelling of on-layer potentials in various TLG polytypes. These calculations make use of the hybrid $k\cdot p$ theory -- tight-binding model (HkpTB) Hamiltonians that take into account both closest and next neighbour hoppings within and between the layers. The influence of magnetic field is taken into account by computing the Landau level (LL) spectra, also with the full self-consistent analysis in the Hartree approximation. \textcolor{black}{This analysis is presented for both low and high temperature (up to room temperature) and taking into account a perturbation due to a substrate.}

\section{Hybrid $k\cdot p$ - tight-binding (H$kp$TB) model of tetralayer graphenes}

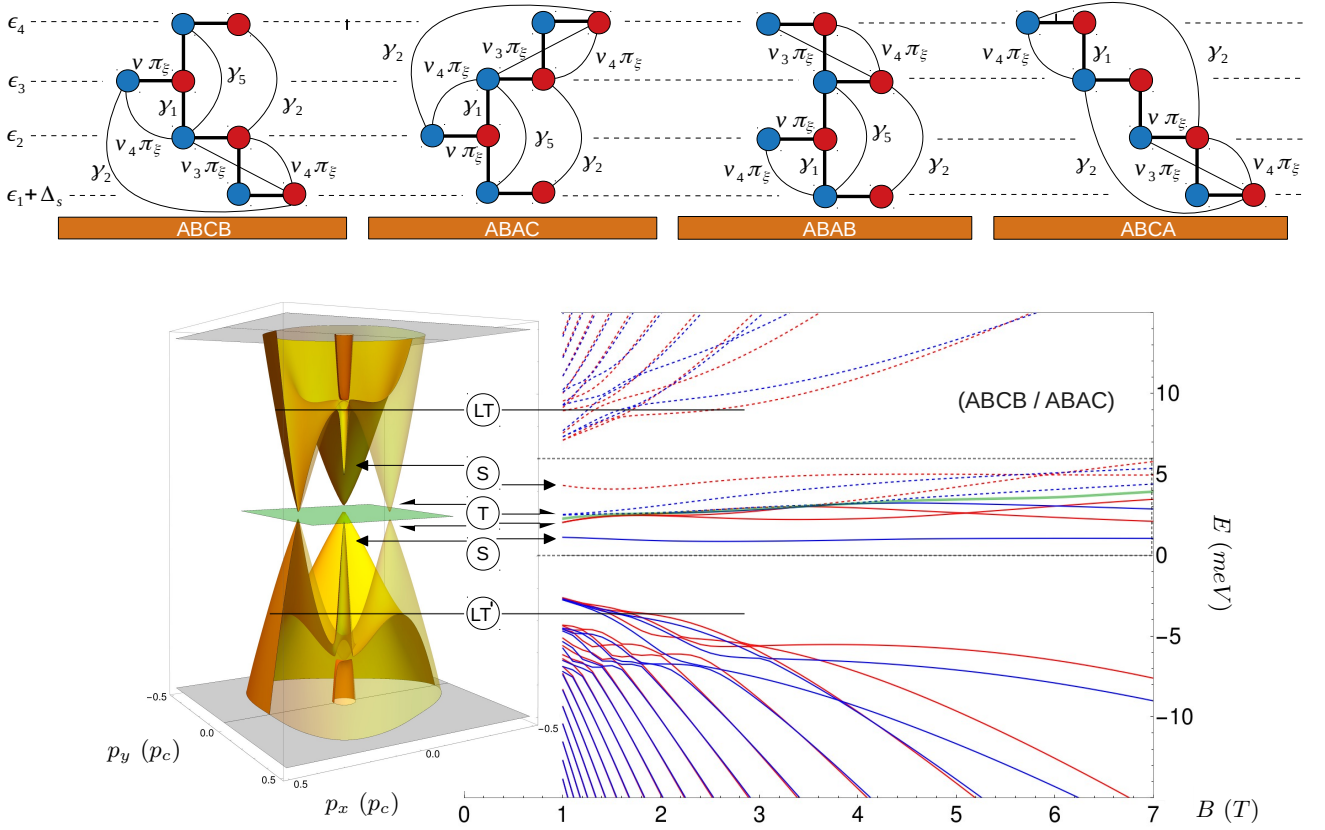
\begin{figure*}[]
\centering
\begin{tikzpicture}
  \node (img1)
  { \includegraphics[width=1\textwidth]{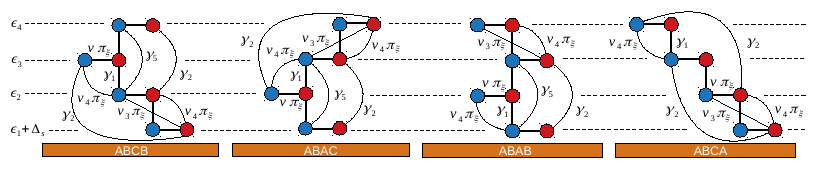}};
  \node (img1)[distance=0cm,xshift=0cm, yshift=-5.8cm]
  { \includegraphics[width=0.8\textwidth]{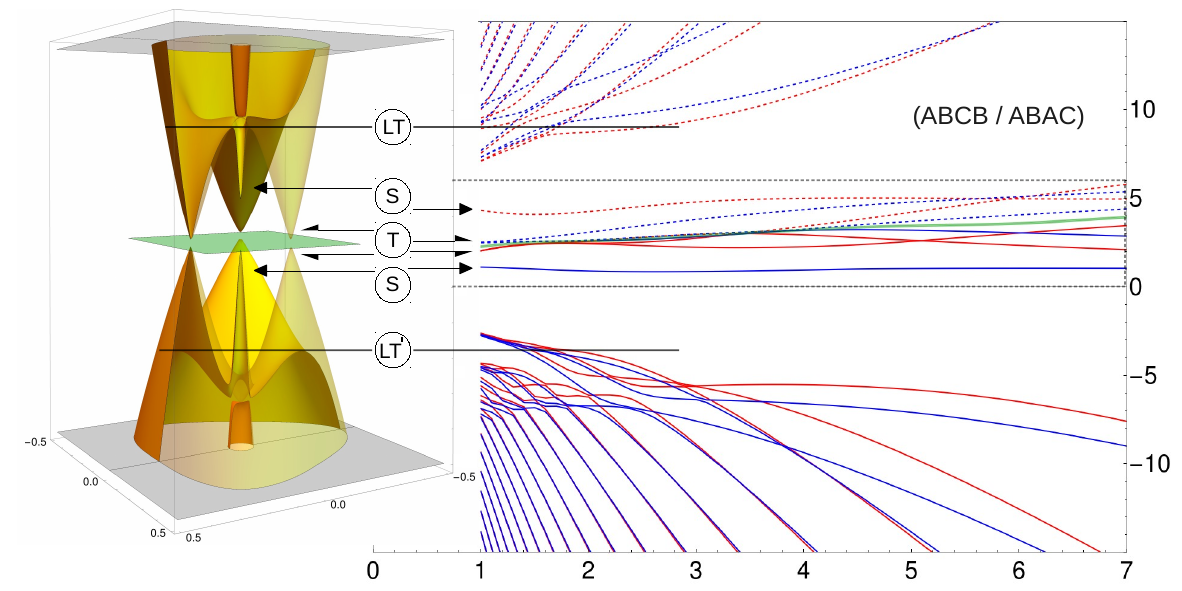}};

  \node[distance=0cm,xshift=7.5cm, yshift=-9.1cm,font=\color{black}] {$B$ $(T)$};
\node[distance=0cm, rotate=-90, anchor=center,xshift=5.8cm,yshift=7.4cm,font=\color{black}] {$E$ $(meV)$};
\node[distance=0cm,xshift=-4.0cm, yshift=-9.0cm,font=\color{black}] {$p_{x}$ $(p_{c})$};
\node[distance=0cm,xshift=-6.9cm, yshift=-8.3cm,font=\color{black}] {$p_{y}$ $(p_{c})$};

\end{tikzpicture}
\caption{The top panels sketch the four possible stacking arrangements of graphene tetralayers. In each structure, we present sublattice A and B as blue and red circles, respectively, coupled by the SWMcC parameters. We also label the on-site energies $\epsilon_{\cal{l}}$ and the substrate-induced energy shift $\Delta_s$, represented by the underlying orange rectangle. To note, positive sign of electric polarisation $P_z$ corresponds to bottom-to-top orientation. In the bottom left panel we show an exemplary band structure of ABCB/ABAC graphene, with $p_x$ and $p_y$ given in units of $p_c=\gamma_1/v$ and with $\Delta_s=0$. The green plane indicates the Fermi energy at charge neutrality. On the bottom right, we present the LL spectrum, with the green line showing the Fermi energy, below (above) which we present the spectrum in solid (dashed) lines. The LLs from the $K_+$ and $K_-$ valley are in red and blue, respectively. The lines labeled LT and LT$^\prime$ represent the position of the two Lifshitz transitions in the non-magnetic spectrum. To note, the 8 spectrally isolated LLs, labelled as triplets (T) or singlets (S), largely determine the change in the electron occupation on each layer, and is shown in greater resolution in Fig. \ref{Fig:LLs_nDiff_Pz}. \label{Fig:BS_LLS}}

\end{figure*}

All four distinguishable stacking configurations of tetralayers (ABCB, ABAC, ABAB, ABCA) are sketched in Fig. \ref{Fig:BS_LLS}, where we also stress that the bottom layer in the film is affected by a substrate. The HkpTB Hamiltonians describing these tetralayers have the following form \cite{mcellistrim_spectroscopic_2023,garcia-ruiz_flat_2023}:

\begin{widetext}

\begin{equation} \label{Eq:ABCB}
H_{ABCB}=
\begin{pmatrix}
H_g+\frac{\Delta^{\prime}}{2}(1 +\sigma_z)+\epsilon_1 +\Delta_{s}&V_\mathrm{AB}&W_{ABC}&0\\
V_\mathrm{AB}^{\dag}&H_g+\Delta^{\prime} +\epsilon_2&V_\mathrm{AB}&W_{ABA}\\
W_{ABC}^{\dag}&V_\mathrm{AB}^{\dag}&H_g+\Delta^{\prime}(1 -\sigma_z)+\epsilon_3&V_\mathrm{AB}^{\dag}\\
0&W_{ABA}^{\dag}&V_\mathrm{AB}&H_g+\frac{\Delta^{\prime}}{2}(1 +\sigma_z) +\epsilon_4
\end{pmatrix},
\end{equation}

\begin{equation} \label{Eq:ABAC}
H_{ABAC}=
\begin{pmatrix}
H_g+\frac{\Delta^{\prime}}{2}(1 +\sigma_z)+\epsilon_1 +\Delta_{s}&V_\mathrm{AB}&W_{ABA}&0\\
V_\mathrm{AB}^{\dag}&H_g+\Delta^{\prime}(1 -\sigma_z)+\epsilon_2&V_\mathrm{AB}^{\dag}&W_{ABC}^{\dag}\\
W_{ABA}^{\dag}&V_\mathrm{AB}&H_g+\Delta^{\prime}+\epsilon_3&V_\mathrm{AB}^{\dag}\\
0&W_{ABC}&V_\mathrm{AB}&H_g+\frac{\Delta^{\prime}}{2}(1 +\sigma_z) +\epsilon_4
\end{pmatrix},
\end{equation}

\begin{equation}  \label{Eq:ABAB}
H_{ABAB}=
\begin{pmatrix}
H_g+\frac{\Delta^{\prime}}{2}(1 +\sigma_z)+\epsilon_1 +\Delta_{s}&V_\mathrm{AB}&W_{ABA}&0\\
V_\mathrm{AB}^{\dag}&H_g+\Delta^{\prime}(1 -\sigma_z)+\epsilon_2&V_\mathrm{AB}^{\dag}&W_{BAB}\\
W_{ABA}^{\dag}&V_\mathrm{AB}&H_g+\Delta^{\prime}(1 +\sigma_z) +\epsilon_3&V_\mathrm{AB}\\
0&W_{BAB}^{\dag}&V_\mathrm{AB}^{\dag}&H_g+\frac{\Delta^{\prime}}{2}(1 -\sigma_z) +\epsilon_4
\end{pmatrix},
\end{equation}

\begin{equation}\label{Eq:ABCA}
H_{ABCA}=
\begin{pmatrix}
H_g+\frac{\Delta^{\prime}}{2}(1 +\sigma_z)+\epsilon_1 +\Delta_{s}&V_\mathrm{AB}&W_{ABC}&0\\
V_\mathrm{AB}^{\dag}&H_g+\Delta^{\prime} +\epsilon_2 &V_\mathrm{AB}&W_{ABC}\\
W_{ABC}^{\dag}&V_\mathrm{AB}^{\dag}&H_g+\Delta^{\prime} +\epsilon_3&V_\mathrm{AB}\\
0&W_{ABC}^{\dag}&V_\mathrm{AB}^{\dag}&H_g+\frac{\Delta^{\prime}}{2}(1 -\sigma_z) +\epsilon_4
\end{pmatrix},
\end{equation}

\begin{equation}   \notag
H_{g}=
\begin{pmatrix}
0&v\pi_{\xi}^{\ast}\\
v\pi_{\xi}&0
\end{pmatrix},
\quad 
V_\mathrm{AB}=
\begin{pmatrix}
-v_{4}\pi_{\xi}&\gamma_1\\
-v_{3}\pi_{\xi}^{\ast}&-v_{4}\pi_{\xi}
\end{pmatrix},
\end{equation}

\begin{equation}   \notag
W_{ABA}=
\begin{pmatrix}
\frac{\gamma_5}{2}&0\\
0&\frac{\gamma_2}{2}
\end{pmatrix},
\quad
W_{BAB}=
\begin{pmatrix}
\frac{\gamma_2}{2}&0\\
0&\frac{\gamma_5}{2}
\end{pmatrix},
\quad
W_{ABC}=
\begin{pmatrix}
0&0\\
\frac{\gamma_2}{2}&0
\end{pmatrix}
\end{equation}

\end{widetext}
Here, we use a basis of sublattice Bloch states $\{\phi_{\cal{l}}^\mathrm{A},\phi_{\cal{l}}^\mathrm{B}\}$ on each layer ${\cal{l}}=1,...,4$, and an expansion of intra- and inter-layer hoppings in powers of $\pi_\xi=\xi p_x+ip_y$, where $(p_x,p_y)$ is the momentum measured from the $\mathbf{K}_\xi=\xi 4\pi/3a (1,0)$ point. In the above matrix structures each entry represents a $2\times2$ block and $\sigma_z$ is the third Pauli matrix. These HkpTB Hamiltonians follow the Sloncewski-Weiss-McClure (SWMcC) \cite{Slonczewsi_Band_1958,McClure_Band_1957,McClure_Theory_1960} parameterisation with  $(v,v_3,v_4)=(1,0.1,0.022)10^6 \,\,\unit{m/s}$  and $(\gamma_1,\gamma_2,\gamma_5,\Delta')=(390,-17,38,25) \,\,\unit{meV}$, adopted from Ref. \cite{yin_dimensional_2019,ge_control_2021}. An on-site energy, $\Delta'$, originates from the dimmer bonds between adjacent layers, where each atom receives a $x\Delta'$ shift, where $x$ is the number of $\gamma_1$ couplings to that site (Fig. 1). We also include the influence of a substrate as an additional on-site potential $\Delta_s$ on the bottom layer, ${\cal{l}}=1$, which will be a variable parameter in the following calculations. In Fig.1 (bottom left) we show a representative example of low-energy sub-bands of a mixed stacking TLG computed using such Hamiltonians.

We also use the earlier-developed model \cite{mccann2006asymmetry,koshino2009gate,Slizovskiy_dielectric_2021,garcia-ruiz_mixed-stacking_2023} for a self-consistent calculation of on-layer electrostatic potentials $\epsilon_{\cal{l}}$. That is, we use the fact that electron Bloch states in the $K_+$ and $K_-$ valley are spread over many lattice sites in each layer so that they experience the mean field on-layer potentials, related to the on-layer electron densities, as $\epsilon_{\cal{l}} - \epsilon_{{\cal{l}}-1}$.

\begin{widetext}
\begin{align}\label{Eq:Hartree}
\epsilon_{\cal{l}} - \epsilon_{{\cal{l}}-1}=\frac{e^2 d}{2 \epsilon_0}\left[(n_{\cal{l}}-n_{{\cal{l}}-1})\frac{1+\varepsilon_z}{2\varepsilon_z}+\sum_{j>{\cal{l}}}\frac{n_j}{\varepsilon_z}-\sum_{j^\prime<{\cal{l}}-1}\frac{n_j^\prime}{\varepsilon_z}\right], \quad
    n_{\cal{l}}=
    4\int \frac{d\mathbf{p}}{(2\pi\hbar)^2}
\sum_{\beta}\left[\left(\sum_{\lambda}|\psi_{\beta}^{\lambda,{\cal{l}}}|^2 f(E_{\beta})\right) -\frac{1}{8}   \right].
\end{align}

\end{widetext}
In the above equation, $\varepsilon_z=2.6$ is the out-of-plane dielectric permittivity of graphene, $d=3.35\,\mathrm{\AA}$ the inter-layer distance, 
$\psi_{\beta}^{\lambda,{\cal{l}}}(\mathbf{p})$ denotes the component of the eigenvector at momentum $\mathbf{p}$ of the band $\beta$ on the layer ${\cal{l}}$ and $\lambda=\mathrm{A,B}$ is the sublattice index. All these calculations are performed at a finite temperature, $T$, taken into account by the Fermi factor, $f(E)=(1+e^{E/T})^{-1}$, and, for $B=0$, we take into account four-fold spin-valley degeneracy.

The effect of an out-of-plane magnetic field $B$ can be included using the Luttinger substitution \cite{luttinger_quantum_1956}, where we choose the Landau gauge for vector potential. Then, for both valleys, $K_{\pm}$, we express all valley momentum terms in Hamiltonians 1-4 using raising ($\hat{a}^\dagger$) and descending ($\hat{a}$) operators of the quantum harmonic oscillator, as  
\begin{equation} \notag
    \pi_{+}=\hbar\frac{\sqrt{2}}{l_B} \hat{a},
    \quad\quad
    \pi_{+}^{\dagger}=\hbar\frac{\sqrt{2}}{l_B} \hat{a}^{\dagger},
\end{equation}
\begin{equation} \notag
    \pi_{-}=-\hbar\frac{\sqrt{2}}{l_B} \hat{a}^{\dagger},
    \quad\quad
    \pi_{-}^{\dagger}=-\hbar\frac{\sqrt{2}}{l_B} \hat{a},
\end{equation}
where $l_B=\sqrt{\hbar/eB}$ is magnetic length. To compute the Landau levels, we use the basis $\{\vert n \rangle_{\cal{l}}^\mathrm{A},\vert n \rangle_{\cal{l}}^\mathrm{B}\}$, where $\ket{n}$ is the eigenstate of order $n$ of the harmonic oscillator \cite{Frank_Handbook_2010}, such that $\hat{a}^{\dag}\ket{n}=\sqrt{n+1}\ket{n+1}$ and $\hat{a}\ket{n}=\sqrt{n}\ket{n-1}$. We truncate the series at a level $N_c$ large enough to ensure spectral convergence. As noted in Ref. \cite{zhang_magnetoelectric_2011}, the maximum value of $N_c$ also depends on the sublattice and layer index, to avoid artificially generated zero-energy Landau levels. Then, in the presence of magnetic field, where the combination of time inversion symmetry breaking with inversion symmery breaking by lattice structure and substrate leads to lifted valley degeneracy, we use the following expression for on-layer densities expressed in terms of wave functions $\psi_{\beta,\xi}^{\lambda,{\cal{l}}}$ of each LL, $\beta$,
\begin{align}\label{Eq:n_B} \tag{6}
n_{\cal{l}}=\frac{1}{\pi l_{B}^2}
\sum_{\xi}
\sum_{\beta}
\left[
\left(
\sum_{\lambda}
|\psi_{\beta,\xi}^{\lambda,{\cal{l}}}|^2 f(E_{\beta,\xi})
\right)-\frac{1}{8}
\right].
\end{align}
 In Fig. 1 (bottom right), we show an exemplary LL spectrum for ABCB graphene film with $\Delta_s=0$, where the low-energy LLs are traced to the relevant sub-band edges at $B\rightarrow 0$ and the inversion symmetry breaking in mixed stacking TLGs leads to the inter-valley LL splittings. Using densities in Eqs. (\ref{Eq:Hartree}) and (\ref{Eq:n_B}) we determine electric polarisations for each polytype, 
\begin{equation}
    P_z=ed\sum_{{\cal{l}}=1}^{4}{\cal{l}}n_{\cal{l}},\nonumber
\end{equation}
and compute the difference in electrical polarisation between the two mixed stacking structures $\mathrm{ABAC}$ and $\mathrm{ABCB}$,
\begin{equation}\tag{7}
    \mathcal{P}=P_{z}^{ABAC}-P_{z}^{ABCB},
\end{equation}
to highlight the sensitivity of polarisation to the inversion asymmetry of those two twins. To mention, this would be the quantity analysed in the recent KPFM studies \cite{atri_spontaneous_2023} of TLGs.

\section{Substrate-induced potential and temperature dependence of electric polarisation of TLG twins}

In this section, we analyse the influence of substrate-induced potential affecting the bottom layer of TLG and the dependence of the out-of-plane electrical polarisation on temperature, from cryogenic up to room temperature, (used in the recent experimental studies \cite{atri_spontaneous_2023}). Using the model and approach described in Section II, we calculate the dependence of electrical polarisation in the mixed stacking tetralayers. The polarisations computed across a broad temperature range for both ABAC/ABCB TLGs with $\Delta_s=0$ and $\Delta_s=5$ meV are plotted in Fig. 2(a) with dashed and solid lines respectively. In Fig. 2(b) we show the difference $\mathcal{P}$ between such polarisations, plotting it both as a function of temperature and substrate-induced potential. \textcolor{black}{The data in Fig. 2(b) and Fig. 2(c) indicates that $\mathcal{P}$ varies linearly with $\Delta_s$, up to at least $100$ meV, which is shown explicitly at $300$ K but has also been verified at all other temperatures.} As a consequence, we can separate the intrinsic and extrinsic contributions to polarisation as
\begin{equation}\tag{8}
  \mathcal{P}=\mathcal{P}(T,\Delta_s=0)+\alpha(T) \Delta_s.  
\end{equation}

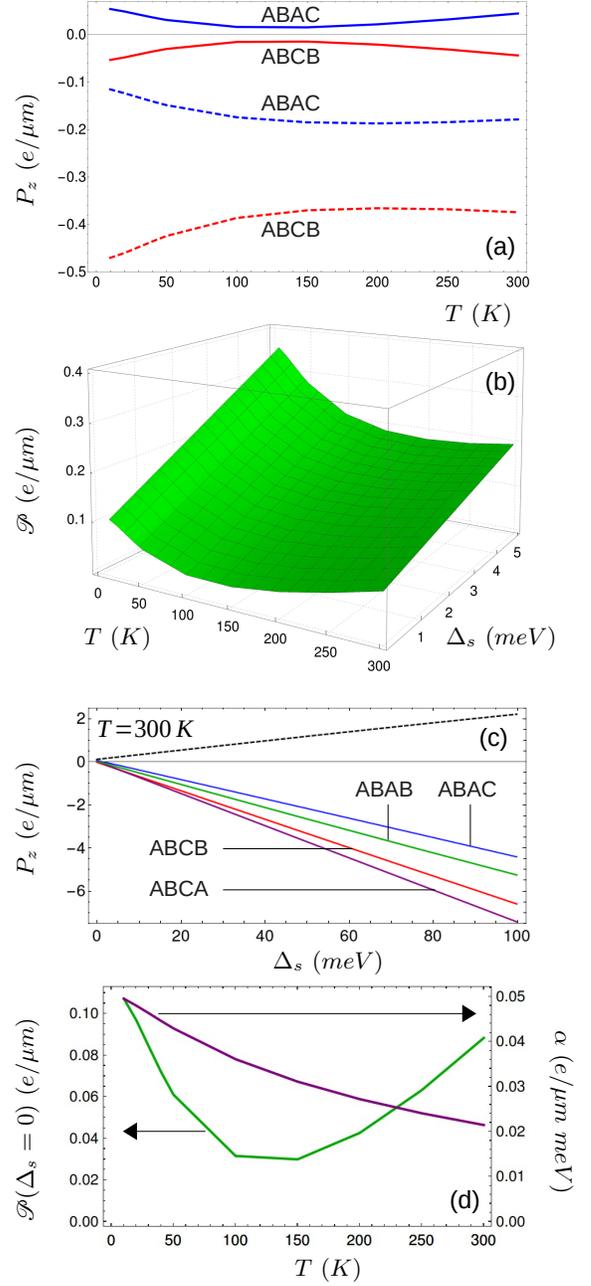
\begin{figure}[H]
\begin{tikzpicture}
  \node (img1)
  { \includegraphics[width=0.45\textwidth]{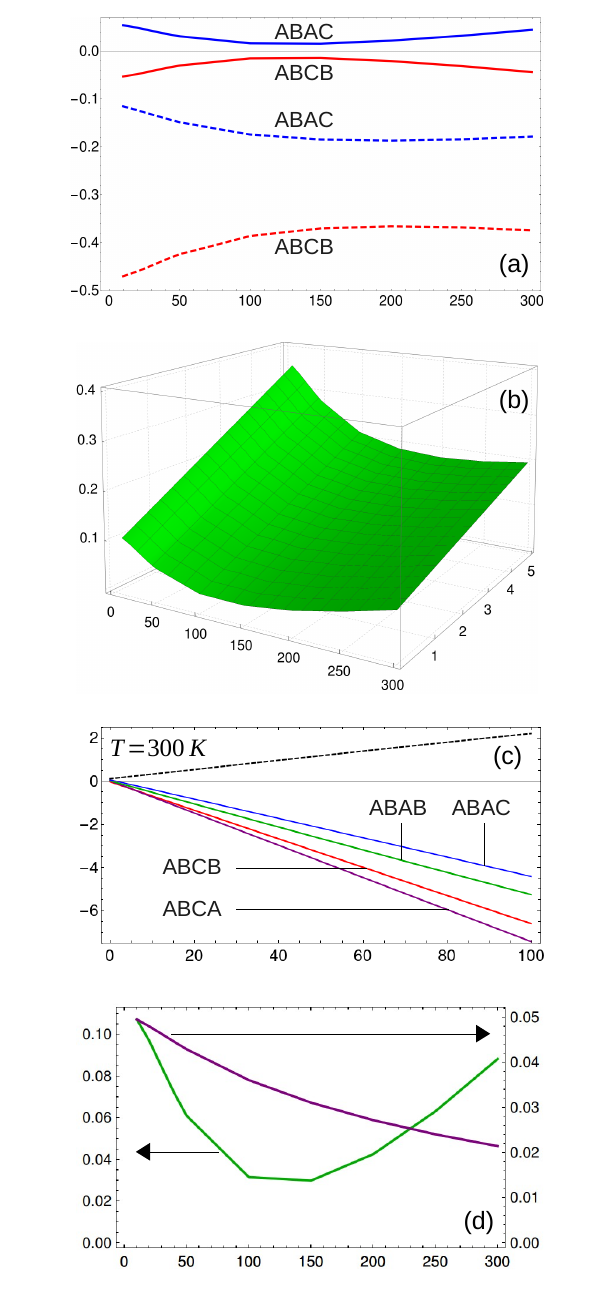}};

  \node[ distance=0cm, rotate=90, anchor=center,xshift=6.4cm,yshift=3.5cm,font=\color{black}] {$P_z$ $(e/\mu m)$};

  \node[distance=0cm, rotate=90, anchor=center,xshift=2cm,yshift=3.5cm, font=\color{black}] {$\mathcal{P}$ $(e/\mu m)$};

   \node[ distance=0cm, rotate=90, anchor=center,xshift=-2.5cm,yshift=3.5cm,font=\color{black}] {$P_z$ $(e/\mu m)$};

  \node[distance=0cm, rotate=90, anchor=center,xshift=-6.5cm,yshift=3.5cm, font=\color{black}] {$\mathcal{P}(\Delta_s=0)$ $(e/\mu m)$};

  \node[distance=0cm, rotate=-90, anchor=center,xshift=6.4cm,yshift=3.6cm, font=\color{black}] {$\alpha$ $(e/\mu m$ $meV)$};

    \node[distance=0cm, rotate=0, anchor=center,xshift=2.8cm,yshift=-0.1cm, font=\color{black}] {$\Delta_s$ $(meV)$};

     \node[distance=0cm, rotate=0, anchor=center,xshift=0.5cm,yshift=-4.41cm, font=\color{black}] {$\Delta_s$ $(meV)$};

    \node[distance=0cm, rotate=0, anchor=center,xshift=-2.3cm,yshift=-0.1cm, font=\color{black}] {$T$ $(K)$};

     \node[distance=0cm, rotate=0, anchor=center,xshift=0.5cm,yshift=-8.5cm, font=\color{black}] {$T$ $(K)$};
     
      \node[distance=0cm, rotate=0, anchor=center,xshift=2.5cm,yshift=4.2cm, font=\color{black}] {$T$ $(K)$};

\end{tikzpicture}
\caption{(a) The $P_z$ temperature dependence of the mixed stacking twins ABCB and ABAC. Solid lines show the intrinsic polarisation with $\Delta_s=0$ meV and dashed lines include a substrate potential $\Delta_s=5$ meV. (b) We plot the difference between ABAC and  ABCB stacking polarisation $\mathcal{P}$ as a function of both temperature T and substrate potential $\Delta_s$. \textcolor{black}{(c) We show that $P_z$ evolves linearly as a function of $\Delta_s$ up to $100$ meV for all stacking configurations at $300$ K. The black dashed line is $\mathcal{P}$ also at $300$ K.} (d) Using the left hand side y-axis we plot $\Delta P_{z}$ with zero substrate $\Delta_s=0$, i.e. the purely intrinsic contribution. Using the right hand side y-axis we plot coefficient $\alpha$ (defined in Eq. 8), also as a function of temperature, illustrating the diminishing contribution of the substrate at greater temperatures.
\label{Fig:T_dep}}
\end{figure}

Here, $\mathcal{P}(T,\Delta_s=0)$ is the intrinsic contribution to $\mathcal{P}$; its computed temperature dependence is plotted in Fig. 2(d) by the green line, and we note that it displays a non-monotonic behavior where an initial decrease at low temperatures is recover upon heating up to room temperature. It is important to note that a finite value of this intrinsic polarisation contribution requires both the lack of inversion symmetry in the lattice structure and an electron-hole asymmetry: as polarisation, which is proportional to the electron charge conjugation would have inverted sign. In terms of the HkpTB Hamiltonians for mixed stacking tetralayers, their minimal version with only the closest neighbour hoppings would have a hidden charge conjugation symmetry \cite{garcia-ruiz_mixed-stacking_2023}, so that the out-of-plane electric polarisation is proportional to the electron-hole symmetry breaking terms $v_4,\Delta^{\prime},\gamma_2,\gamma_5$ in the full SWMcC model. 

In contrast, the extrinsic contribution, accounted by the second term in Eq. (8), is independent of those SWMcC model parameters, and it is linear in $\Delta_s$, with a coefficient $\alpha(T)$. The computed values of $\alpha(T)$ are shown in Fig. \ref{Fig:T_dep}(d) (purple line), which indicates that the size of the extrinsic effect slowly decreases with temperature. In principle, the difference between the trends in temperature dependence of intrinsic and extrinsic contributions to $\mathcal{P}$, in particular, in the temperature range of $\sim 50-300$K can be used to judge the relevance of each of these contributions. For example, if one would observe a decrease in the size of polarisation at higher temperatures, this would be indicative of the dominant role of substrate $\Delta_s$.

We use Eq. (8) with $\mathcal{P}(300K,\Delta_s=0)$ and $\alpha(300K)$, based on the data in Fig. \ref{Fig:T_dep}(d), to estimate the substrate-induced potential $\Delta_s$ that would correspond to the experimental observations made in Ref.\cite{atri_spontaneous_2023}. The TLG flakes studied in those KPFM measurements were exfoliated onto a SiO$_2$ substrate. Using the value of $\mathcal{P}$ measured in Ref.\cite{atri_spontaneous_2023} at room temperature, we estimate that the SiO$_2$ substrate induces $\Delta_s = 75 \pm 5$meV on the bottom graphene layer, and the magnitude of the reported $\mathcal{P}$ is dominated by the substrate effect distinguishing stacking orders of ABCB and ABAC twins. On a positive note, we point out that measurements of polarisation of the mixed stacking twins on a variety of underlying substrates can offer a method for determining the substrate-induced on-layer potentials for graphene. To mention, the value $\Delta_s^{\rm{G/SiO}_2}\sim 70-80$meV determined here for graphene on SiO$_2$ is comparable to the difference, $\Delta_s^{\rm{G/hBN}}-\Delta_s^{\rm{G/G}}\approx 18$meV, between proximity potentials induced by hexagonal boron nitride (hBN) and a strongly misaligned graphene (G), determined \cite{boschi2024builtin} from the measurements of the inter-layer asymmetry gap \cite{mccann2006asymmetry} in bilayer graphene (BLG) in G/BLG/hBN stacks.

\section{Magnetic field effect on TLG polarisation}

The comparison of the measured \cite{atri_spontaneous_2023} and computed values of  polarisation difference between ABCB and ABAC twins, discussed at the end of Section III, shows that there is a need for devolving the substrate effect on the magnitude of $\mathcal{P}$ from its `intrinsic' value. Below, we consider the polarisation dependence on magnetic field to identify features specific to the intrinsic part of $\mathcal{P}$. The idea here is that inversion-asymmetry of ABCB and ABAC twins, together with time-reversal asymmetry due to magnetic field, lifts valley degeneracy of Landau levels (LLs), which may produce some characteristic magnetic field dependence of `intrinsic' $\mathcal{P}$. 

The computed LL spectrum of a `suspended' ABCB TLG, with $\Delta_s =0$, is displayed in Fig. \ref{Fig:BS_LLS}(bottom-right). This LL spectrum has the following specific features:

\begin{itemize}[noitemsep]
\item At low energies $\sim-10<E<10$ meV we identify triads of LLs in $K_{\pm}$ valleys which can be traced to the three mini-valleys in the zero field dispersion on the bottom left hand side of Fig. \ref{Fig:BS_LLS}.
\item These three mini-valleys which give rise to the triad LLs are labeled ``T" in Fig. \ref{Fig:BS_LLS}. They are slightly different in $K_{+}$ and $K_{-}$ valleys due to the broken degeneracy. These LLs are almost non-dispersive and are comparable to the LLs in rhombohedral trilayers \cite{slizovskiy_films_2019}. 
 \item We also identify two additional non-degenerate zero energy (non-dispersive) LLs marked as ``S" in Fig. \ref{Fig:BS_LLS}, which can be traced to the slightly gaped, central Dirac features in the zero field spectrum.
\item The LL spectrum takes more mundane structure (valley degeneracy is less pronounced) outside the energy interval set by van Hove singularities in the spectrum marked by LT (as related to the Lifshitz transition in the band topology 
\cite{varlet_anomalous_2014}).
\item The exemplary LL spectrum in Fig. \ref{Fig:BS_LLS} identifies a group of eight LLs which host the Fermi level in undoped ABCB/ABAC graphenes, shown by the green line. Due to the broken valley degeneracy and splitting of mini-valley triplets there are multiple crossings of LLs from different valleys in the magnetic field range $\sim 1.5 - 4.5$T where LL proximity is small. When LLs inter-cross near the Fermi-energy the occupation of LLs change as a function of magnetic field in accordance with Pauli principle.
\end{itemize}

To mention, for each value of magnetic field, these LL spectra were calculated taking into account self consistency between on-layer potentials and change distribution between the layers for an overall neutral TLG.

\begin{figure}[]
\begin{tikzpicture}
  \node (img1)
  { \includegraphics[width=0.45\textwidth]{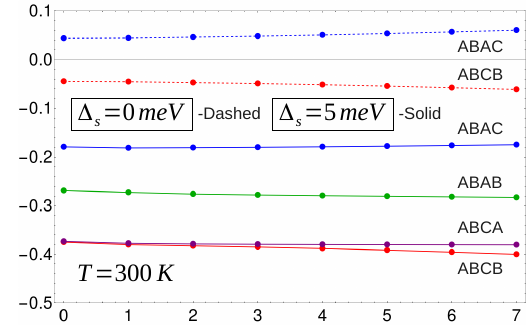}};

  \node[distance=0cm, rotate=90, anchor=center,xshift=0cm,yshift=4.2cm, font=\color{black}] {$P_{z}$ $(e/\mu m)$};

      \node[distance=0cm,xshift=2.4cm, yshift=-2.7cm,font=\color{black}] {$B$ $(T)$};
  
\end{tikzpicture}
\caption{Electric polarisation of of TLG at room temperature with a substrate potential of $\Delta_s=0$ (dashed) for the mixed stacking twins and $\Delta_s=5$ meV (solid) for all four polytypes. For $\Delta_s=0$, ABAB and ABCA polarisations are absent. At high temperatures, it is almost impossible to distinguish between ABCB and ABCA graphenes without a strong magnetic field.
\label{Fig:B_dep}}
\end{figure}

The magnetic field dependence of polarisation at room temperature ($300$K) for all four charge neutral TLG allotropes is shown in Fig. 3. Dashed lines are for $\Delta_s=0$, the intrinsic polarisation in mixed stacking twins. For inversion-symmetric Bernal and rhombohedral stackings, their symmetry forbids any out-of-plane polarization both at $B=0$ and finite magnetic fields. Solid lines have been computed for a substrate-induced asymmetry potential, $\Delta_s=5$ meV. Due to the inversion symmetry breaking by the substrate, our self-consistent calculation of out-of-plane spontaneous electric polarisation returns non-zero effects for all four structures. We see a monotonic trend in magnetic field with little variation due to high temperature smearing of the Fermi-step ($\sim26$meV) in comparison to the inter-LL spacing seen in Fig. \ref{Fig:BS_LLS} ($\sim1$meV). Since each LL has different corresponding charge distribution, the overall contribution to $P_z$ is shared amongst many LLs near the Fermi energy such that the fine ordering of these levels become unimportant at such high temperatures. At all strengths of magnetic field the magnitude of $P_z$ is primarily determined by the size of $\Delta_s$.

Additionally, we notice that the polarisation of ABCA and ABCB graphenes are almost indistinguishable without the addition of a strong magnetic field. They are also most strongly affected by the substrate compared to the other stacking orders. This is due to the surface states present in rhombohedral graphene which result in additional surface bound charge on the layer, in direct contact with the substrate. ABCB graphene, despite having mixed stacking order, has its rhombohedral part on the substrate side of the structure, therefore yielding very similar response to the substrate as is seen in the pure rhombohedral stack.

In contrast to the high temperature results, at low temperatures (few-Kelvin range) we see the emergence of a non-monotonic trend with magnetic field. First we look at the intrinsic electric polarisation in charge-neutral tetralayers by setting $\Delta_s=0$ in equations 1-4. 
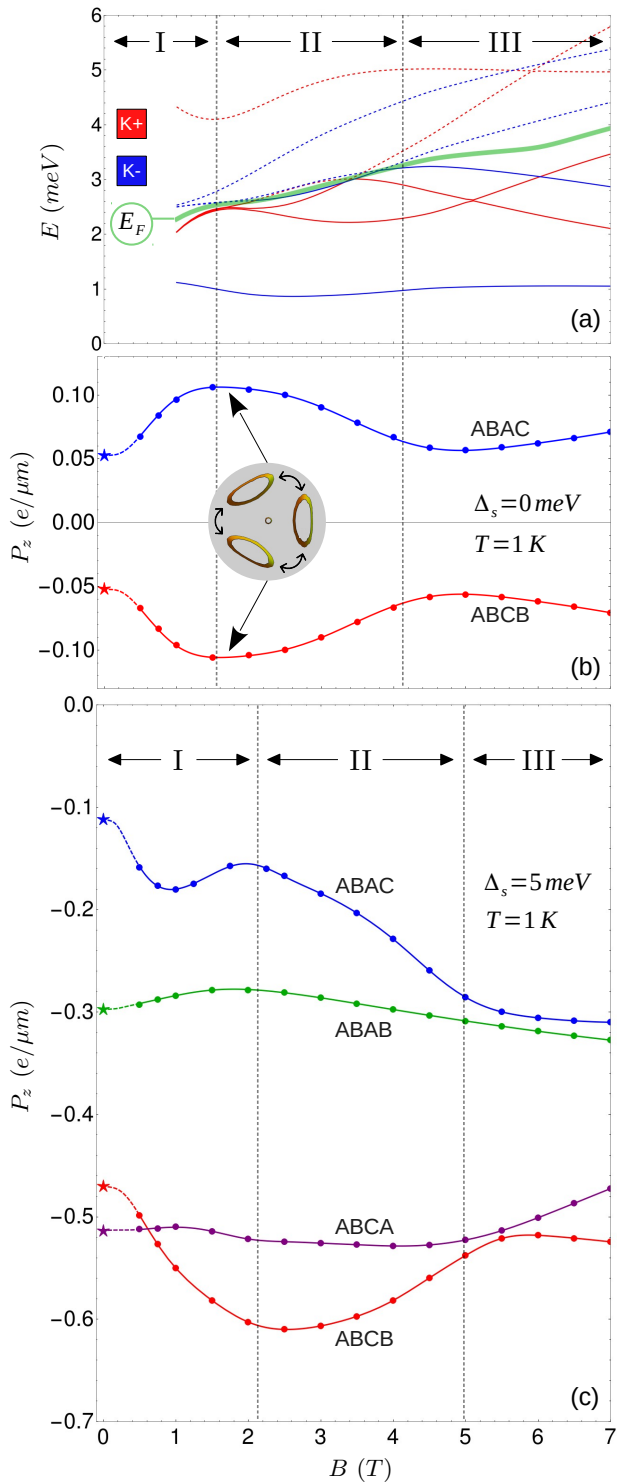
\begin{figure}[H]
\begin{tikzpicture}
  \node (img1)
  { \includegraphics[width=0.45\textwidth]{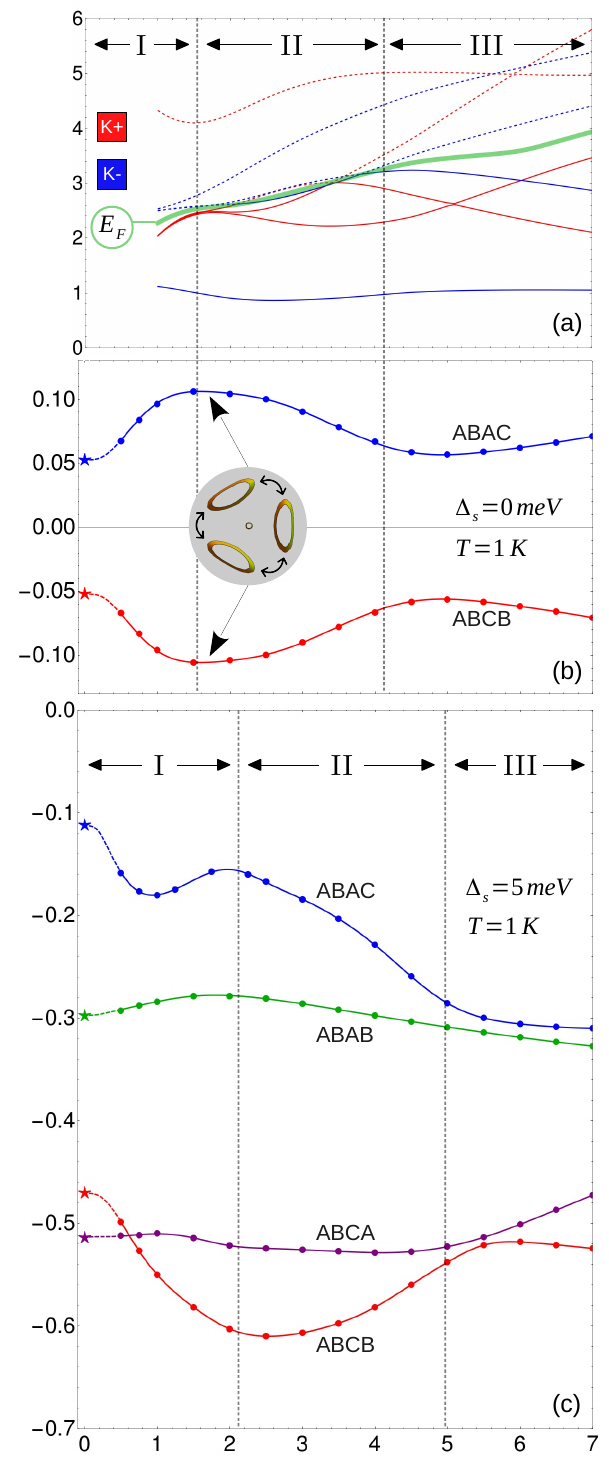}}; 
\node[distance=0cm,xshift=0.5cm, yshift=-9.8cm,font=\color{black}] {$B$ $(T)$};
\node[distance=0cm, rotate=90, anchor=center,xshift=7.2cm,yshift=3.6cm, font=\color{black}] {$E$ $(meV)$};
\node[distance=0cm, rotate=90, anchor=center,xshift=3cm,yshift=4cm,font=\color{black}] {$P_z$ $(e/\mu m)$};
\node[distance=0cm, rotate=90, anchor=center,xshift=-4.3cm,yshift=4cm,font=\color{black}] {$P_z$ $(e/\mu m)$};

\end{tikzpicture}
\caption{(a) Spectrum of the zero-th order Landau levels in ABCB and ABAC graphene, with $\Delta_s=0$, for the $K_+$ and $K_-$ valleys. Green line represents the Fermi level with width $k_B T$  broadening for $T =1K$. (b) Out-of-plane electric polarisation of ABCB and ABAC graphene. We sketch a slice of the low energy band structure, indicating the mixing of states due to magnetic field. This is the magnetic break down effect which causes the splitting of triplet LLs. (c) Out-of-plane electric polarisation now with $\Delta_s=5$ for all polytypes.
\label{Fig:LLs_nDiff_Pz}}
\end{figure}
In Fig. \ref{Fig:LLs_nDiff_Pz} (a), we zoom in to the spectrum of those eight non-dispersive LLs identified previously in Fig. \ref{Fig:BS_LLS}, to follow valley and mini-valley splitting effects and the resulting crossing of LLs originating from different valleys. This allows us to identify three characteristic magnetic field intervals: \\
{\bf I}. Lowest magnetic fields where there are valley split mini-valley triplets which lead to a valley polarised state.\\
{\bf II}. Intermediate interval where the splitting of triplets, changes the order of LLs in $K_+$ and $K_-$ valleys. This interval includes two LL crossing points. The beginning of this interval is approximately where magnetic break down occurs, which mixes the three mini-valleys.\\
{\bf III}. High magnetic field range where, despite any further crossings, there are two completely filled LLs which originate from the split triplet in the $K_+$ valley. There is also two completely filled LLs from the $K_-$ valley, one from the $K_-$ triplet and one from the $K_-$ singlet (the other four from this eight level group are empty), resulting in a net zero valley polarisation of the system.

Overall, we find a non-monotonic behavior of $P_z(B)$ at low temperatures, with reversed trends in ABCB and ABAC twins and characteristic for the vicinity of the above-mentioned LL crossings, marked on Fig. 4(b). This non-monotonic dependence is a feature of intrinsically induced polarisation, which also appears as `bulges' in the substrate-perturbed structures, Fig. 4(c) (where we compare the results obtained for all four TLG polytypes, with the corresponding LL spectra displayed in Appendix (\ref{appendix:Bigfig}). Therefore, we suggest that, by identifying and tracing these non-monotonic features, one could distinguish the intrinsic contribution to $P_z$ caused by stacking order from the substrate effects. In this respect, we note that in the self-consistent  analysis of the present study was limited to the Hartree approximation, whereas a full Hartree-Fock theory \cite{jung_lattice_2011} could reveal the formation of various spin-valley polarised quantum Hall effect liquids forming at the lowest temperatures at the fields where the LL crossings appear in Fig. 4(a), which we leave for future studies.

Finally, Fig. 5 compiles the information obtained in this work about the parametric (temperature, magnetic field, and substrate proximity potential) dependence of ABCB/ABAC TLG polarisation, encoded in $\mathcal{P}=\mathcal{P}(T,B,\Delta_s=0)+\alpha(T,B) \Delta_s$, at higher temperatures spanning $20K-300K$ which may be directly used for the interpretation of further KPFM studies of such systems. 

\begin{figure}[]
\begin{tikzpicture}
  \node (img1)
  { \includegraphics[width=0.45\textwidth]{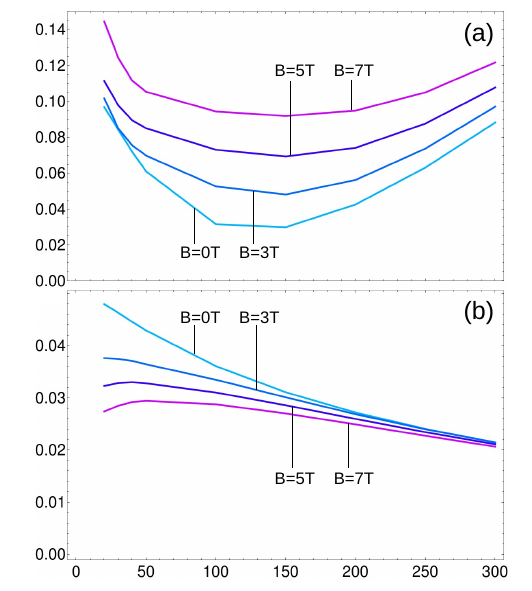}};

  \node[distance=0cm, rotate=90, anchor=center,xshift=2cm,yshift=3.9cm, font=\color{black}] {$\mathcal{P}(\Delta_s=0)$ $(e/\mu m)$};

   \node[distance=0cm, rotate=90, anchor=center,xshift=-2.1cm,yshift=3.9cm, font=\color{black}] {$\alpha$ $(e/\mu m$ $meV)$};

     \node[distance=0cm, rotate=0, anchor=center,xshift=0.9cm,yshift=-4.7cm, font=\color{black}] {$T$ $(K)$};

\end{tikzpicture}
\caption{(a) We show how the intrinsic $\mathcal{P}$ ($\Delta_s=0$) evolves as a function of temperature for different strengths of magnetic field. (b) Plotted, is the coefficient of extrinsic contribution to $\mathcal{P}$, $\alpha$ (defined in Eq. (8)), again, as a function of temperature for different strengths of magnetic field. The contribution from the substrate is reduced as temperature is increased and does not change much with magnetic field at high temperatures. In contrast, at low temperatures, magnetic field can diminish the substrate effect almost as much as temperature can.
\label{Fig:Substrate_Surface}}
\end{figure}
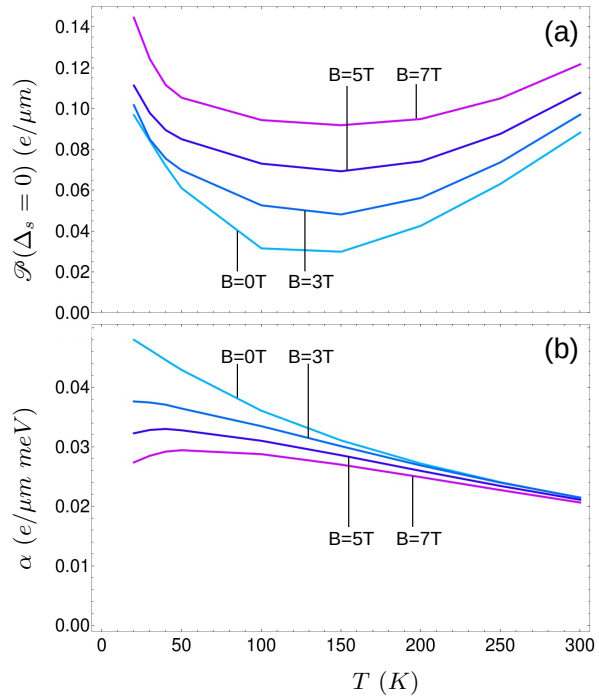

\section{Discussion and Conclusions}

Overall, we analysed the out-of-plane electric polarisation of four polytypes of tetralayer graphene (ABAB Bernal, ABCA rhombohedral, and ABCB/ABAC mixed stacking configurations), taking into account both the influence of a substrate (as a `proximity' potential, $\Delta_s$ induced on the bottom layer) and the intrinsic contribution due to the lack of inversion symmetry in ABCB and ABAC twins. The computed (using the HkpTB approach) polarisation values and their dependence on temperature and external magnetic field, $B$, indicate that, even when comparing the polarisation of twins, $\mathcal{P}=\mathcal{P}(T,B,\Delta_s=0)+\alpha(T,B) \Delta_s$, the substrate effect can easily overtake the intrinsic contributions. 
Such a sensitivity to the `proximity' potential, induced by the substrate on the bottom graphene layer, can be traced to the surface states characteristic to the rhombohedral part of the mixed-stacking polytype. Having applied the computed polarisation dependence to the analysis of the substrate induced potential measured in Ref.\cite{atri_spontaneous_2023} by KPFM of TLGs on a SiO$_2$ substrate, we concluded that those earlier-observed values were dominated by the substrate effect, and estimated $\Delta_s^{\rm{G/SiO}_2}\sim 70-80$meV for graphene on SiO$_2$. 

\begin{figure}[]
\begin{tikzpicture}
  \node (img1)
  { \includegraphics[width=0.45\textwidth]{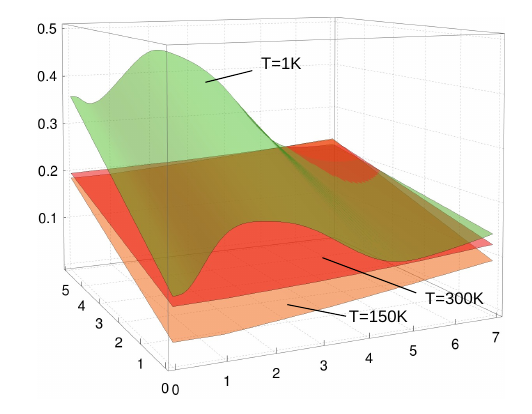}};

  \node[distance=0cm, rotate=90, anchor=center,xshift=1.2cm,yshift=3.9cm, font=\color{black}] {$\mathcal{P}$ $(e/\mu m)$};

      \node[distance=0cm,xshift=3cm, yshift=-2.7cm,font=\color{black}] {$B$ $(T)$};
  
   \node[distance=0cm, rotate=0, anchor=center,xshift=-3.3cm,yshift=-2.3cm, font=\color{black}] {$\Delta_s$ $(meV)$};
    
\end{tikzpicture}
\caption{$\mathcal{P}$, defined in Eq. (7) as a function of magnetic field strength and substrate potential at $1K$ in green, $150K$ in orange and $300K$ in red. At low temperature we see the three regime trend as before. This non-monatonic trend is lost at higher temperatures due to broadening of the Fermi step becoming much larger than LL separation. We see changes in $\mathcal{P}$ with $\Delta_s$, due to greater charge accumulation on layer 1 of ABCB graphene compared with layer 1 of ABAC graphene, with the effect of the substrate felt more by the former. One sees that for greater magnetic field strength, the effect of changing $\Delta_s$ is progressively suppressed as the surface flattens. 
\label{Fig:Substrate_Surface}}
\end{figure}
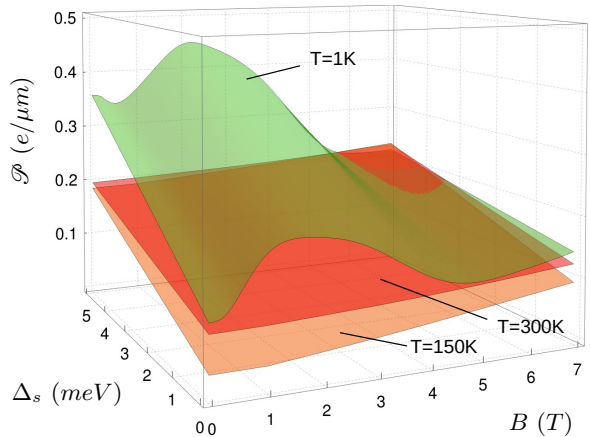

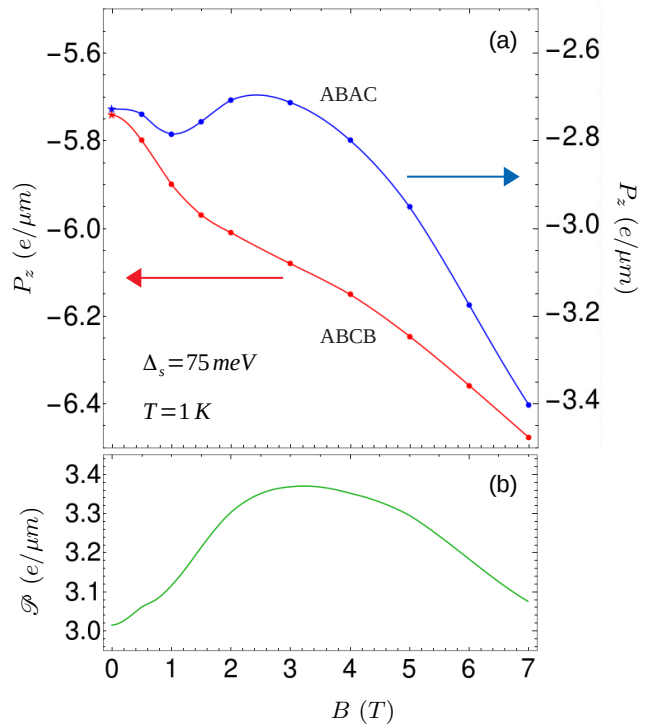
\begin{figure}[]
\begin{tikzpicture}
  \node (img1)
  { \includegraphics[width=0.43\textwidth]{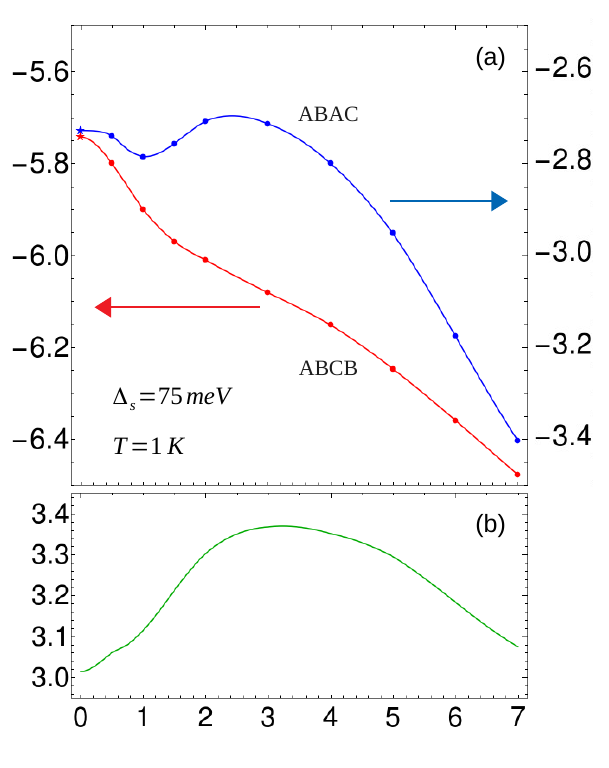}}; 

 \node[distance=0cm, rotate=90, anchor=center,xshift=1.5cm,yshift=4cm, font=\color{black}] {$P_z$ $(e/\mu m)$};

  \node[distance=0cm, rotate=-90, anchor=center,xshift=-1.5cm,yshift=4cm, font=\color{black}] {$P_z$ $(e/\mu m)$};

  \node[distance=0cm, rotate=90, anchor=center,xshift=-2.8cm,yshift=3.9cm, font=\color{black}] {$\mathcal{P}$ $(e/\mu m)$};

      \node[distance=0cm,xshift=0.5cm, yshift=-4.8cm,font=\color{black}] {$B$ $(T)$};

\end{tikzpicture}
\caption{\textcolor{black}{(a) The low temperature ($1$K) magnetic field dependence of $P_z$ for ABAC and ABCB graphene using the large predicted $\Delta_s^{\rm{G/SiO}_2}= 75$ meV. (b) The difference of the two curves shown in (a), $\mathcal{P}$. Highlighting the continued presence of the non-monotonic bulge, even at such a high proximity induced substrate potential.}
\label{Fig:Substrate_Surface}}
\end{figure}

We also note that the intrinsic contribution to the polarisation may be highlighted at low temperatures and quantising magnetic fields, due to the valley splitting of the TLG Landau levels caused by the interplay between time-inversion symmetry breaking by magnetic field and lack of inversion symmetry in the mixed-stacking twins. The specific feature of the intrinsic polarisation contribution appears in Fig. 6 as a bulge in the magnetic field dependence of $\mathcal{P}$, which develops at low temperature and signals the inter-valley LL crossings, but does not appear at high temperatures where the individual LL occupancy is not resolved due to the thermal smearing of the Fermi step. 

\textcolor{black}{For completeness, we analyse the low temperature ($1$K) magnetic field dependence of the mixed stacking TLGs $P_z$ with our estimated value of $\Delta_s^{\rm{G/SiO}_2}= 75$ meV in Fig. 7(a). The non-monotonic feature observed for ABCB graphene has been suppressed, as the band edges from which the crossing triplet LLs emerge are pushed apart by the large substrate potential, resulting in less LL crossing. In contrast the non-monotonic feature is prominent for ABAC graphene as the aforementioned band edges are shifted in the same direction and remain in relatively close proximity such that the resulting LL triplets cross paths. Consequently, the non-monotonic bulge in $\mathcal{P}$ predicted above is ever present, and even enhanced slightly in magnitude by the strong proximity induced substrate potential as shown in Fig. 7(b).}

\section*{Acknowledgements}
We thank Moshe Ben Shalom, Niels Walet and Chistian Moulsdale for useful discussions.

This work was supported by EC-FET Core 3 European Graphene Flagship Project, EPSRC grants EP/S030719/1 and EP/V007033/1, the National Science and Technology Council (NSTC 112-2112-M-006-019-MY3) and the Lloyd Register Foundation Nanotechnology Grant. P.J.S was supported by Graphene-NOWNANO Center for Doctoral Training.

\appendix
\section{Landau levels for substrate perturbed structures}\label{appendix:Bigfig}

\begin{figure*}[t!]
\centering
\begin{tikzpicture}
  \node (img1)
  { \includegraphics[width=0.95\textwidth]{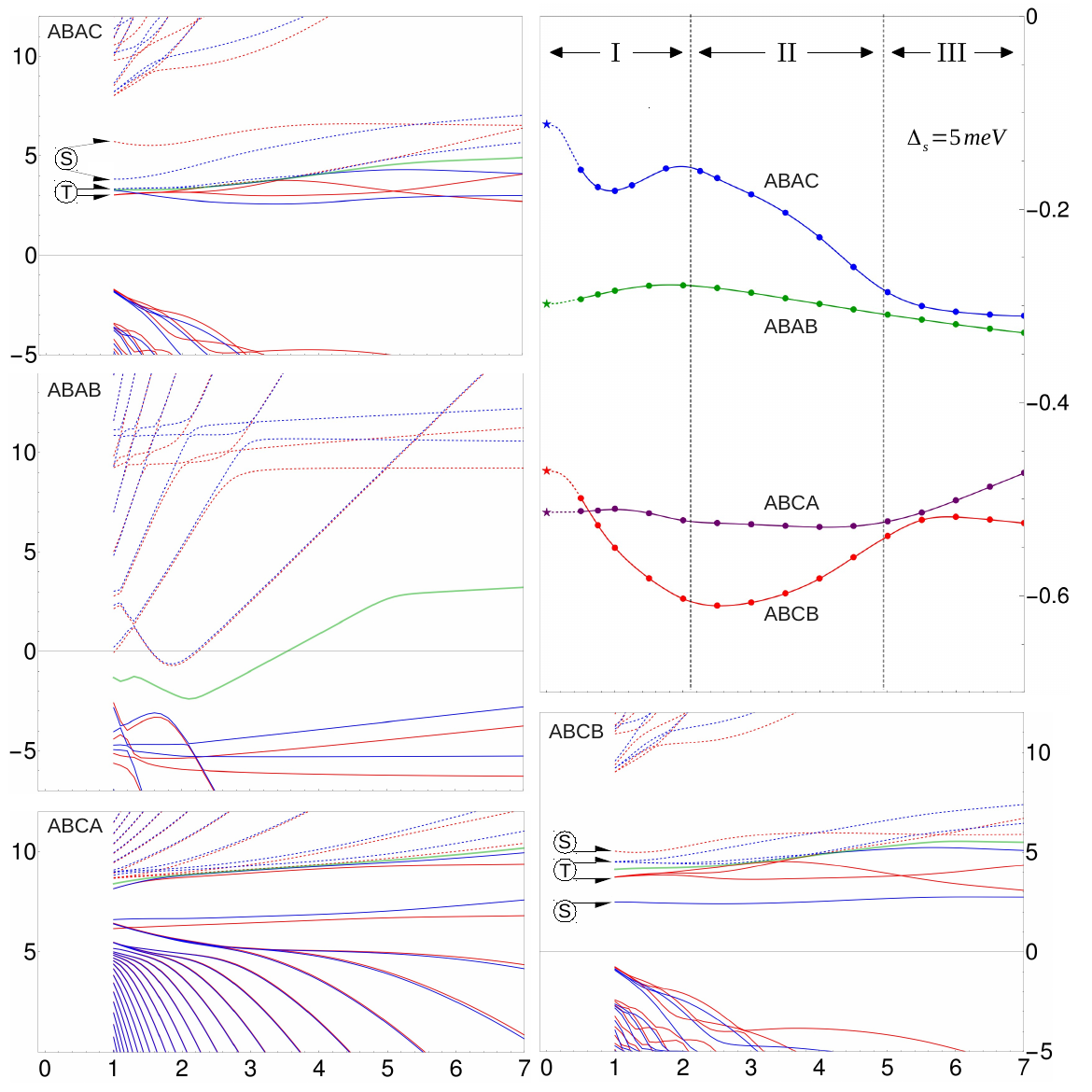}};
  \node[distance=0cm,xshift=-4cm, yshift=-8.65cm,font=\color{black}] {$B$ $(T)$};

  \node[distance=0cm,xshift=3.85cm, yshift=-8.65cm,font=\color{black}] {$B$ $(T)$};
  
  \node[distance=0cm, rotate=90, anchor=center,xshift=5.5cm,yshift=8.55cm,font=\color{black}] {$E$ $(meV)$};
  
   \node[distance=0cm, rotate=90, anchor=center,xshift=-1.2cm,yshift=8.55cm,font=\color{black}] {$E$ $(meV)$};
   
    \node[distance=0cm, rotate=90, anchor=center,xshift=-6cm,yshift=8.55cm,font=\color{black}] {$E$ $(meV)$};

    \node[ distance=0cm, rotate=-90, anchor=center,xshift=5.4cm,yshift=8.15cm,font=\color{black}] {$E$ $(meV)$};

     \node[ distance=0cm, rotate=-90, anchor=center,xshift=-3.65cm,yshift=8.2cm,font=\color{black}] {$P_z$ $(e/\mu m)$};
\end{tikzpicture}
\caption{The top right panel shows electric polarisation of all four TLG stacking orders (shown in Fig. 1) as a function of magnetic field strength with a substrate potential $\Delta_s=5$ meV. We calculate $P_z$ for $B=0$ shown by the star markers for each structure, and interpolate between the data with and without B-field using a dashed parabola. All corresponding LLs are shown in the remaining four panels. We also identify the three regimes in electric polarisation and corresponding regimes of differing LL occupation. Colour coding, solid/dashed LLs and the Green stripes have the same meaning as described in Fig. 2 and Fig. 3. We once again Identify the position of LL triplets (T) and singlets (S) for the mixed stacking TLGs.\label{Fig:Substrate_Effects}}
\end{figure*}
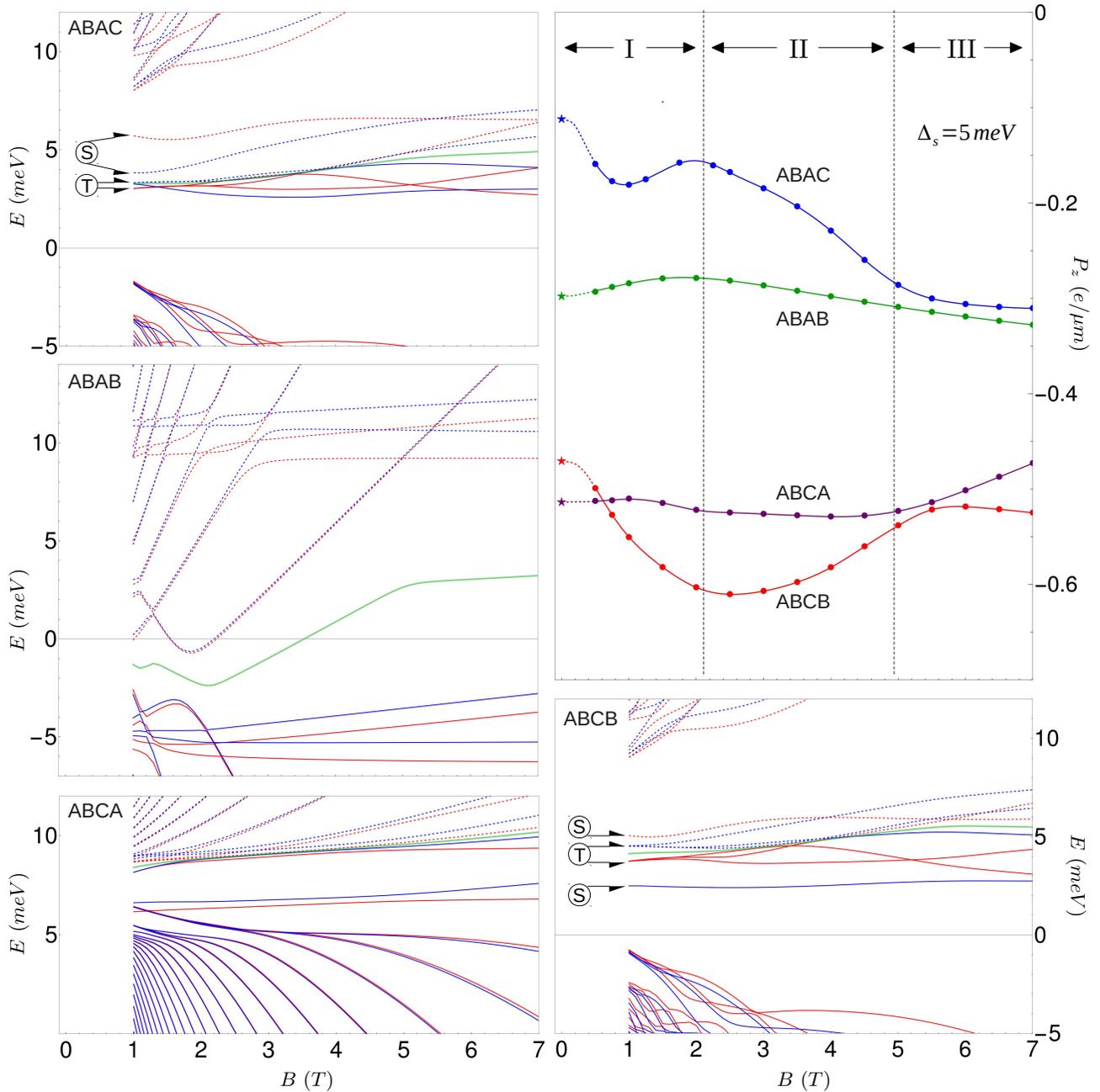

In Fig. \ref{Fig:Substrate_Effects}, we present the LL spectra for all four TLG allotropes computed for a substrate-induced asymmetry potential, $\Delta_s=5$ meV at $1$K shown in the top-right panel. The polarisation data in the panel also include the values at $B=0$ which differ for ABAC and ABCB by much more than those found without taking substrate into account. The reason for this is that these two twin structures are differently affected by the substrate, proportionally to $\Delta_s$. Nevertheless, $P_z(B)$ dependence in both of those twins retain the non-monotonic ``hump like" features identified for $\Delta_s=0$, and at high magnetic fields their difference becomes approximately the same as in the interval III identified in Fig. \ref{Fig:LLs_nDiff_Pz} for a TLG without a substrate. 

Further details of magnetic field dependence of $P_z$ can be understood by applying a magnifying glass to the LL spectra in Fig. \ref{Fig:Substrate_Effects}. For example, if we inspect the LLs of the ABAC tetralayer we notice that the singlet zero energy LLs are never filled in either of the two valleys across the studied magnetic field range, where as in ABCB TLG the $K_-$ ``S" state is always below the Fermi level this indicates that the highlighted features in the magnetic field dependence originate from the LL triplets, which originate from the mini-valley band edges whose splitting is less affected by the substrate.

For the inversion-symmetric configurations, where the source of charge asymmetry is entirely due to the substrate, we find that the value of the polarisation for any magnetic field strength is larger for the rhombohedral-stacked configuration. This can be understood in terms of its low-energy dispersion, with states highly localised around the outermost layers, where the excess in electron charge induces the largest change on $P_z$. The trend in $P_z$ is not monotonic. This is due to the close proximity and resultant mixing of LLs in regime II, leading to differing LL occupation between the three regimes despite the lack of LL crossing points in the bottom left panel of Fig. \ref{Fig:Substrate_Effects}. Only Bernal configuration exhibits a purely monotonically (past 1.5 T) decreasing value of $P_z$ as a function of magnetic field. This is due to the clear spectral isolation of the LLs above and below the Fermi energy, which at $1$ K are fully depleted and fully occupied, respectively. Charge distribution responds linearly to the magnetic field applied as occupation of LLs change linearly with magnetic field.

\bibliographystyle{unsrt}

\bibliography{References}

\end{document}